\documentclass{notices}
\usepackage{amsmath}
\usepackage{amsfonts}
\DeclareMathOperator{\tr}{Tr}
\def\fref#1{Fig.~\ref{#1}}
\def\eref#1{Eq.~(\ref{#1})}
\def\tref#1{Thm.~\ref{#1}}
\def\sref#1{Section~\ref{#1}}

\def\M{\mathrm{M}}
\def\T{\mathrm{P}}
\def\W{\mathrm{W}}
\def\AO{{\A\Omega}}
\def\A{\mathrm{A}}
\def\M{\mathrm{M}}
\def\AO{{\A\Omega}}
\def\AM{\mathrm{AM}}

\def\T{{\mathrm{P}}}

\usepackage{mhequ}
\usepackage{tikz}
\usepackage{wrapfig}
\usepackage{siunitx}
\usetikzlibrary{tikzmark}
\newcommand{\tikzarc}[1]{%
\tikzmarknode{a}{#1}
\begin{tikzpicture}[overlay,remember picture]
\draw ([yshift=1pt]a.north west) to[bend left=20] ([yshift=1pt]a.north east);
\end{tikzpicture}%
}
\def\sph#1{\tikzarc{#1}} 
\def\sAO{{\sph{\AO}}}
\def\sAM{{\sph{\AM}}}

\def\sT{{\sph{{\T}}}}
\def\nb{\mathbf{n}}
\def\rb{\mathbf{r}}
\def\rc{\rb_{\rm c}}
\def\rshell{r_{\rm shell}}
\def\l{\left}
\def\r{\right}

\def\d{\mathrm{d}}

\let\phi=\varphi
\let\epsilon=\varepsilon

\usepackage{amsfonts,amssymb,amsmath,amscd,amsthm}
\newtheorem{definition}{Definition}
\newtheorem{conjecture}{Conjecture}
\newtheorem{theorem}{Theorem}
\def\real{{\mathbb{R}}}

\usepackage{tikzsymbols}

 \title{Tumbling Downhill along a Given Curve}

\author{  Jean-Pierre Eckmann
  \affil{is a retired professor of
  Theoretical Physics and Mathematics at the University of Geneva,
  Geneva, Switzerland. His email is jean-pierre.eckmann@unige.ch
  }
\and
  Yaroslav I. Sobolev
  \affil{is a senior research fellow at the Center for Soft and Living
    Matter, Institute for Basic Science (IBS), Ulsan, Republic of
    Korea. His email is yaroslav.sobolev@gmail.com
  }
  \and
  Tsvi Tlusty
    \affil{is a professor at the Center for
  Soft and Living Matter, Institute for Basic Science (IBS) and Department of Physics, Ulsan National Institute of Science and Technology (UNIST), Ulsan, Republic of Korea. His email
  is tsvitlusty@gmail.com}
  }

\begin{document}
\maketitle

\section{The problem}
A cylinder will roll down an inclined plane in a straight line. A cone
will roll around a circle on that plane and then will stop rolling.
We ask the inverse question: For which curves drawn on the inclined
plane $\real^2$ can one carve a shape that will roll downhill
following precisely this prescribed curve \emph{and} its
translationally repeated copies? See \fref{fig:rollingprince} for an example.

This simple question has a solution essentially always, but it turns out that for most curves, the shape will return to its initial orientation only after crossing a few copies of the curve - most often \emph{two} copies will suffice, but some curves require an arbitrarily large number of copies.


\section{Rolling Stones} There is an ample mathematical literature on rolling objects, but for our purpose, it may suffice to mention the study of rolling acrobatic apparatus \cite{segerman} and the rolling of balls on balls or Riemann surfaces \cite{levi}. ``Rolling'' here means pure rolling motion without slipping or pivoting (that is, rotating around an axis perpendicular to the surface). 
With these rules of the game, a cylindrical stone will roll down along a straight line, indefinitely, in the projected direction of gravity. 

But besides straight lines traced by cylinders, can one chisel objects that will follow more interesting planar paths -- curved, convoluted, and self-crossing?
This problem was posed and amply discussed in \cite{xxx}, and the aim of the current contribution is to go into more mathematical detail and add some new results about the general set of paths for which solutions exist.

\begin{figure*}[ht]
        \centering 
        \includegraphics[angle=0,width=\textwidth]{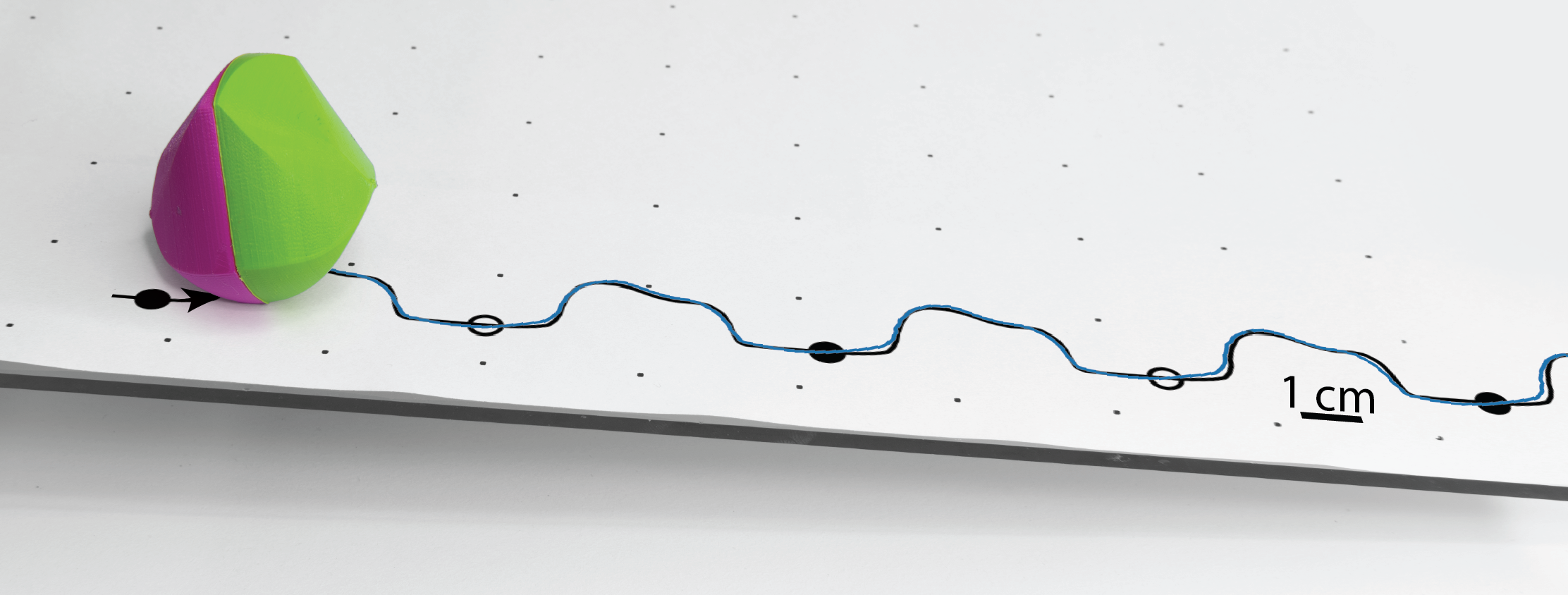}
        \caption{The performance of an actual trajectoid rolling downhill (left-to-right). To demonstrate that trajectoids can be fabricated for rather general paths, we take as a prescribed path  (black curve) the shape of \textit{``a boa constrictor digesting an elephant''}, often mistaken to be simply \textit{``a hat''}, from \emph{The Little Prince} \cite{prince}, which is repeated periodically.
        The blue curve is followed by the actual fabricated trajectoid shown on the left, which has outer diameter of \SI{4.128}{cm}. Black in-plane 1-cm scale bar is shown on the right. Deviations from the ideal black curve are attributed to the precision of the 3D printing, some inertia, and the errors of determining the location of the center of mass (CM) by the projection centroid method \cite{xxx}.
        Note that this trajectoid is made of \textit{two} identical pieces (green and pink), and returns to its original orientation after rolling along \textit{two} periods (this double periodicity is denoted by alternating solid and empty black circles along the flat path).
        Adapted with permission from Fig.~4j of \cite{xxx}.}
\label{fig:rollingprince}
\end{figure*}

A solution means that we have a way to sculpt such a wobbly stone, which we call `trajectoid' because it has the following property: Once placed at the starting point of the trajectory, appropriately oriented, and released, the trajectoid stone should roll along a pre-described infinite trajectory, and \emph{only} along it. This periodic path is made by concatenating translated copies of an original finite path  $\T$, as shown in \fref{fig:rollingprince}.

The principle of trajectoid design (whose realization in practice we will discuss later) is the following: We take a heavy ball and coat it with some lightweight material, which we assume is weightless.
Thus, the rolling object is inhomogeneous, and we will see that this allows the existence of trajectoids.
The featherlight solid envelope is precisely chiseled (or ``shaved''), such that it will lift the heavy ball whenever it tries to depart from the prescribed path. Gravity will therefore ensure that the object's center of mass (CM) always stays at a constant height above the plane, and as a result, this trajectoid obediently follows the desired path.
The concentration of all the mass in the heavy ball allows us define ``following'' simply as
\begin{definition}
\label{def:following}
    A trajectoid \textbf{follows} a given path if a perpendicular projection of the ball's center of mass  traces the path, exactly.
\end{definition}

We typically design and build such gadgets for rolling downhill on an
inclined plane (as in \fref{fig:rollingprince}). But from a
mathematical point of view, it makes more sense to think about a
slightly more general scenario where one puts the trajectoid on a
horizontal plane into a certain starting position and orientation.
One then starts rolling it carefully by hand, without ever sliding, or
rotating the trajectoid around the axis perpendicular to the plane
(``pivoting''), or ever lifting its CM, as in \fref{fig:two_shaves}.  In this scenario, ``uphill'' or self-crossing paths are allowed and feasible simply by changing the direction of the rolling hand.

We will discuss the mathematics of this problem, and note that the question is somewhat different from what is seen in oloids \cite{oloid}, balls rolling on balls or Riemann surfaces \cite{levi}, or a general discussion of acrobatic rolling apparatus \cite{segerman}.
In these cases, the objects are relatively simple, while for our inverse problem, each given path requires its own adapted wobbly stone, and complicated paths require extremely elaborate chiseling of the stone. 

\section{The shaving solution}

To see how one can design
a rolling stone as was done in \cite{xxx}, we consider first a simple  polygonal curve $\T$. The reader then might think of many other curves that are well-approximated by polygons with infinitesimal segments. We now repeat this curve periodically by adding successively identical copies to its end while maintaining the overall orientation in $\real^2$.
More precisely, if $\A$ and $\Omega$ are beginning and end of the path $\T$ then we consider
$\T_{\infty} = \bigcup_{m\in\mathbb{Z}} \l(\T + m\overrightarrow{\A \Omega}\r)$. We want to construct an object that will roll indefinitely along this infinite curve, either on an inclined plane, using gravity, or when we roll it by hand with the constraints described above. 
In our opinion, the beauty of this problem lies in it being a quite simple question with a relatively deep answer.

\begin{figure}[t!]
        \centering 
        \includegraphics[width= 0.45 \columnwidth]{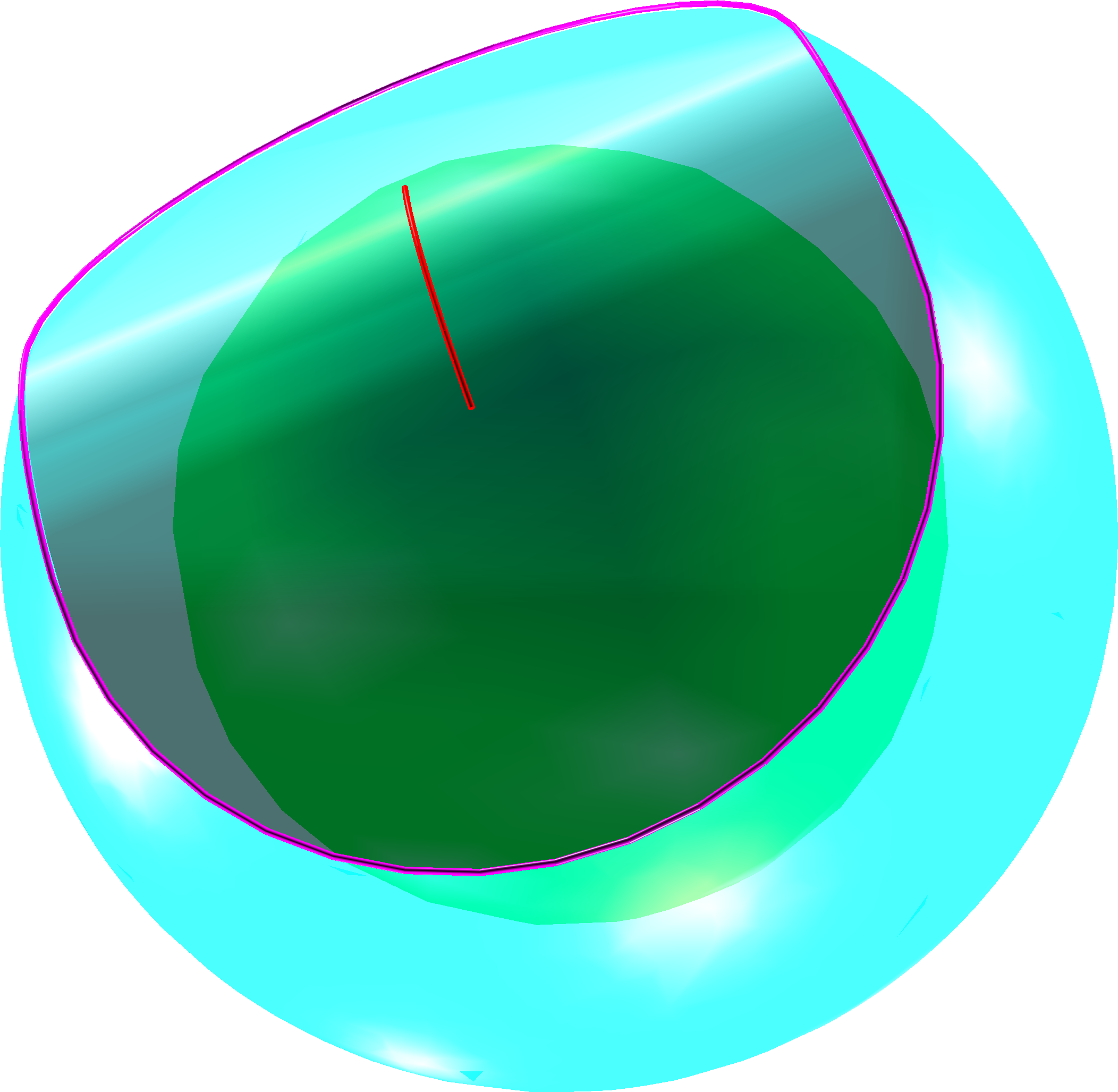}\qquad
        \includegraphics[width= 0.45 \columnwidth]{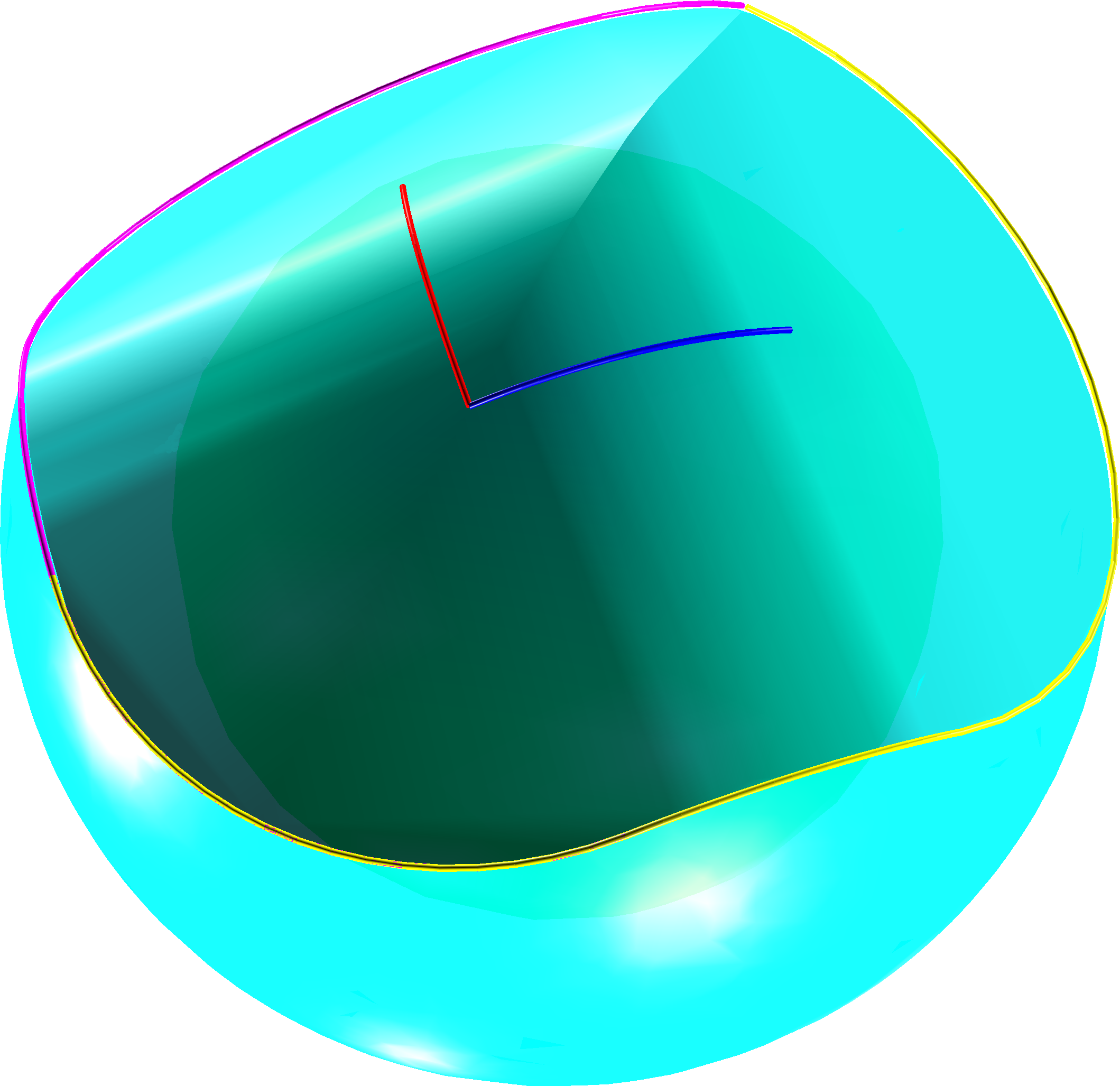}\vskip0.5cm
           \centering    \includegraphics[width=\columnwidth]{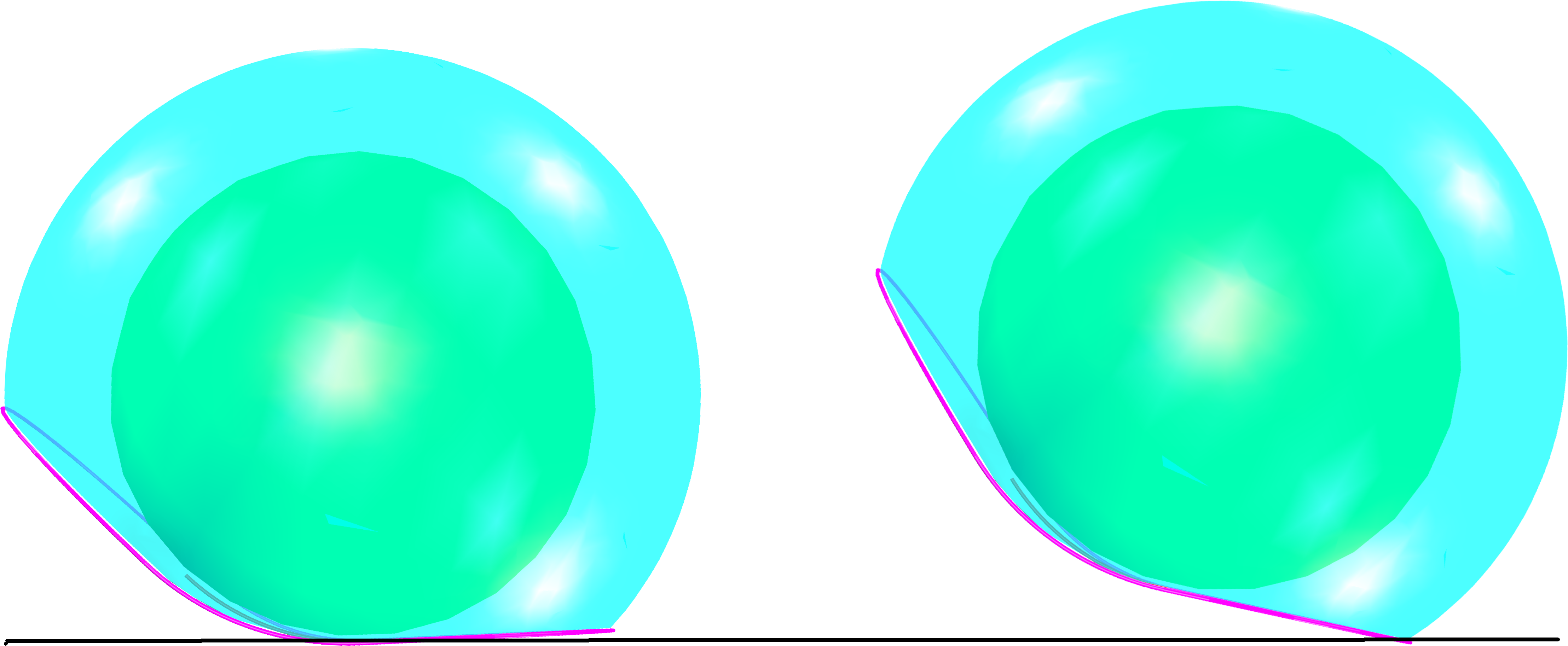}
        \caption{{\bf{Top}}: The shaving.\\  Left: A green ball of radius \SI{1}{\cm}
          surrounded by a light blue ball of radius \SI{1.4}{\cm}. A
          cylindrical surface is shaved from the blue envelope, across
          the violet line. The resulting object can roll along the red
          curve, which is a great arc on the inner green ball, while
          keeping its center of mass (CM) at fixed height (\SI{1}{\cm}, the
          inner ball's radius). Like a cylinder, the object will
          roll along a straight line, but its CM will be
          elevated as soon as it would try to roll beyond one of the
          ends of the red curve.\\ 
        Right: 
        After a second shaving, the object will turn and roll along the blue line, again keeping the CM at the same constant height of \SI{1}{\cm}.\\
{\bf Bottom}: The effect of rolling beyond the cut.\\Left: The position when the object reaches the end of the red arc.\\ 
        Right: The CM of the object is lifted as soon as
        it rolls on beyond this point, and such a motion is therefore
        prevented by gravity. Similarly, tilting sideways also lifts
        the CM.}
\label{fig:two_shaves}
\end{figure}

Our starting point is the very simple observation that a cylinder rolls along any piece of a straight line, with the axis of the cylinder perpendicular to that segment. Hence, by intersecting two cylinders, we can construct an object that rolls along two consecutive straight segments, as illustrated in \fref{fig:two_shaves}. 

We now want to generalize this idea so we can follow 
a polygonal path with many segments. This path $\T=\AO$ is parameterized by its arc-length $t \in [0,L]$, where $L$ is the overall length of $\T$. For our purpose, it is essential to note that:
\begin{enumerate}
    \item The planar path will always touch the inner, heavy ball,
      while the specifically-shaped weightless shell serves to
      stabilize the motion. We take the radius of the heavy ball to be
      $r$, and the maximum radius of the weightless shell to be $\rshell$.
    \item Therefore, by the definition of ``following'' (Def. \ref{def:following}), the rolling motion can be decomposed into the pure translation of the heavy ball's CM, at a height $r$ above the prescribed path $\T$, and a pure rotation of the heavy ball around its CM.
\end{enumerate}

This allows us to formalize the shaving procedure by following the rolling motion in the frame of reference of the moving CM (now located at the origin $\rb = (0,0,0)$). In these coordinates, rolling is a pure rotation around the CM, and the current contact point is always at $\rb = (0,0,-r)$. As the object keeps rolling, the contact point leaves a trace $\rc(t)$ on the surface of the inner ball. Let us examine the shaving procedure for just one straight segment in the plane. 
This corresponds to an arc on the ball, as shown in
\fref{fig:two_shaves}. At each point $\rc(t)$ along the arc, we shave
away any portion of the envelope shell lying beyond the plane tangent
to the inner ball at $\rc(t)$, which serves as the ``razor.''
For each point $\rc(t)$ of the curve, let $ C \l(\rc(t)\r)$ be the
\emph{cut at} $\rc(t)$ This is the half space
\begin{equ}
\label{eq:trajectoid}
C \l(\rc(t)\r) \equiv \l\{\rb: ~~   \rb \cdot \rc(t) \le r^2 
     ~\bigcap~  |\rb| \le \rshell \r \} ~.
\end{equ}
The complement of $C(\rc(t))$ is shaved away. Thus, after traversing
the whole arc, the shaved piece is
\begin{equa}
  C(\mathrm{arc}) = \bigcap_{t\in\mathrm{arc}}  C\l(\rb_{\rm{c}}(t)\r)~.
\end{equa}
We could have shaved away $\{r: |r|\le \rc\}$ all at once at the end
of the procedure, rather than at each step. But this corresponds
better to the mechanical view of \fref{fig:two_shaves}.

Neglecting the mass of the outer shell and the effects of inertia
(or equivalently, inclining the plane so gently that the gadget rolls very slowly), this construction ensures that the object will roll stably and precisely over the arc. The object can follow a general polygon by similarly shaving along each segment, and this holds for any curve $\T$ which can be approximated by finer and finer polygons. The computed object will follow $\T$, but only \emph{once}, and will not necessarily follow its repetitions. 


We note that shaved surfaces of the trajectoid are locally cylindrical (having zero Gaussian curvature), and therefore can be isometrically flattened (or ``developed'') onto a plane without any stretching, compressing, shearing, or tearing. It is easy to see from \eref{eq:trajectoid} that when the outer shell is thick enough (large $\rshell/r$ ratio), the whole outer boundary ($|\rb|=\rshell$) is shaven off, such that the whole trajectoid surface is ``developable.'' Hence, if a trajectoid solution exists for a given path, then by enlarging the outer shell, we can always obtain a \emph{developable trajectoid} for this path. Such trajectoids belong to the class of developable rollers, which includes sphericons, polycons and platonicons \cite{SeatonHirsch2020}. 

\section{The rotation group}

Up to this point, no deep mathematics is needed for the construction. But we would like to achieve more: Namely, if we repeat the original drawn path $\T$ indefinitely, is there still a solution?
Clearly, we need that---after having run over the first repetition of the path---the gadget should be in exactly the \emph{same} orientation as at the starting point.  In more
technical jargon, one then says that the holonomy is a pure translation.  

In \fref{fig:rollingprince}, we show a path and the trajectoid which was fabricated to follow this particular path ``indefinitely.'' The figure shows just four repetitions of the path, with the prescribed path shown in black and the actually followed path in blue, demonstrating the reachable quality of making the gadget by 3D-printing, as explained in \sref{sec:details}.

The question we now ask is when such a construction is possible, and here, some deeper mathematics comes in. Before we go into details, one should note the scaling involved in the
problem. Namely, if one has a solution, then by scaling the curve and the object by the same factor, one again has a solution. In the discussion below, we will therefore fix the curve and only adapt the radius $r$ of the ball until a certain mathematical condition is satisfied. If there exists such a radius $r$, we say that a trajectoid exists for the given curve, and if there is no such $r$, the curve has no solution.
\vskip 0.5cm
\centerline{\emph{Which curves have solutions?}}
\medskip

Running over a polygon connecting $\A$ to $\Omega$, the cumulative effect of rolling is just the product of a sequence of rotations of the shaved ball:
$$
R_{A\Omega}= R_n\cdots R_2 R_1~,
$$
when there are $n$ segments in the polygon, with each rotation $R_j\in \text{SO}(3)$.
To require that the orientation of the piece at the endpoint $\Omega $ of the trajectory returns to its initial orientation at the starting point $\A$ is the same as requiring that the rotation product is $R_{\A\Omega}=\mathbf{1}$, the identity.

 \begin{figure*}[t]
        \centering 
        \includegraphics[width=0.95\textwidth]{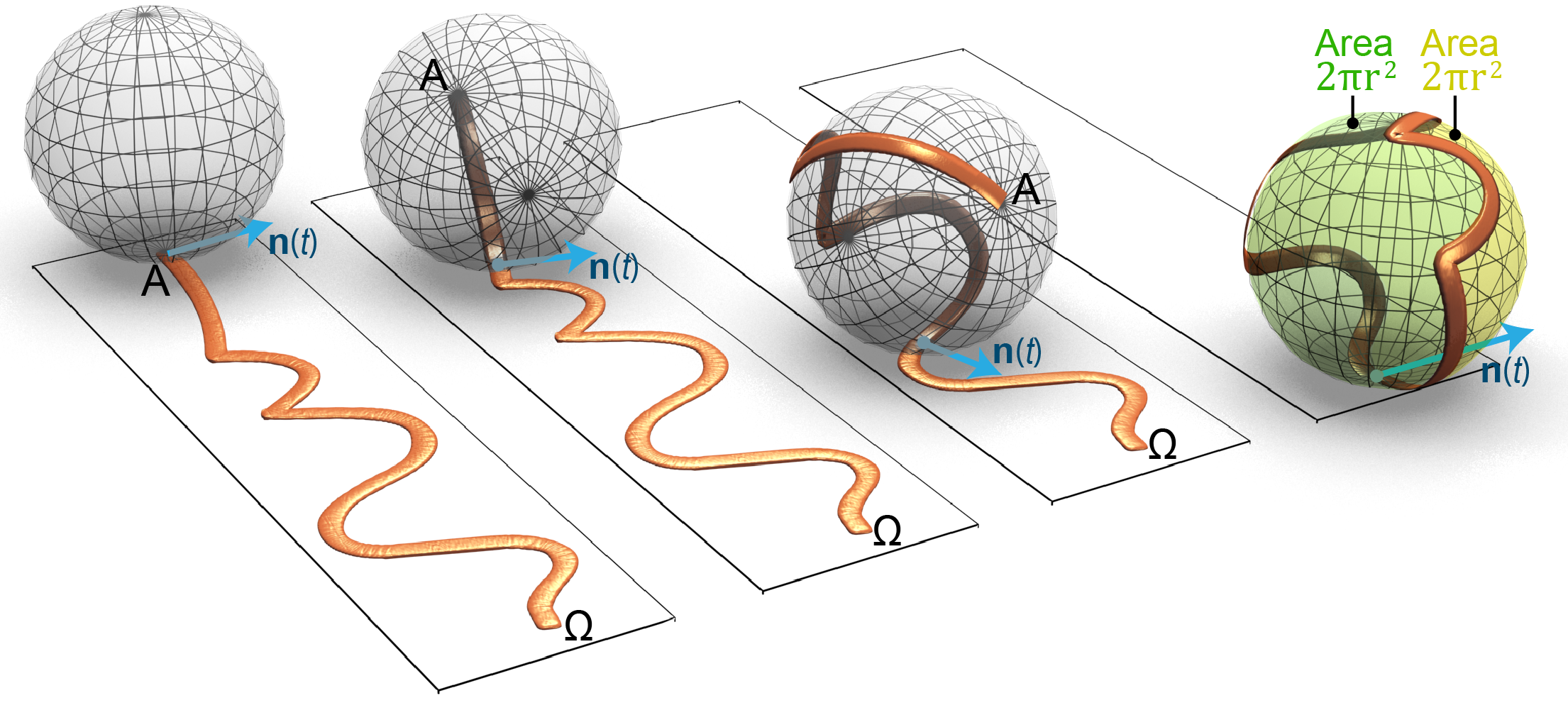}
         \caption{When the ball rolls
           downhill, the path
           $\T = \AO$
           is mapped isometrically onto the ball as $\overset{\frown}{\T}$.
         In particular, the geodesic curvature of the planar path,
         $\kappa(t)$, is conserved. 
        The sufficient condition for a trajectoid to exist is that
        this curve encloses half of the ball's surface (the
        rightmost stage).  
        Adapted with permission from Fig.~2a of \cite{xxx}.}
\label{fig:half_area}
 \end{figure*}

In \fref{fig:half_area}, we illustrate how a given path is actually mapped
from the plane onto the ball. We denote the mapped objects by an overarc $\sph{\phantom{AA}}$~.
The condition $R_{A\Omega}=\mathbf{1}$ means that, when mapped
onto the ball, the initial point $\A$ meets the final point $\Omega$, namely, the curve $\sT={{\sAO}}$ on the ball (as in \fref{fig:half_area}) must be \emph{closed}.

But we also need a condition to guarantee that the orientation of the ball is the same at $\A$ and at $\Omega$. For this, we need to understand the rotation on the surface of the ball, which involves the use of an index:
Consider a path $\T=\AO$, parameterized by its arc length $t \in
[0,L]$, where $L$ is the length of $\T$. Then we can describe $\T$ by
the normal $\nb(t) =  (\cos{ \psi(t)}, \sin{\psi(t)},0)\in\real^3$ to
the path
(blue in \fref{fig:half_area}) in the plane on which the ball rolls,
and $\psi$ is the angle this normal forms (in $x,y$ coordinates on the plane).
Equivalently, we can specify the path using its in-plane (geodesic) curvature $\kappa(t) = \d \psi/\d t$, when the initial angle $\psi(0)$ is given. 

We consider for the path $\T$ the integral over the curvature,
\begin{equation}
\label{eq:index}
I_\T = \frac{1}{2\pi}\int_\T {\kappa(t) \d {t}} = \frac{\Delta \psi}{2\pi} ~,
\end{equation}
where we integrate over one period of $\T$.

When we map $\T$ onto the ball (\fref{fig:half_area}), as
${\sT}={\sAO}$, then, by isometry, the integral is also defined on the
ball, and is equal to the integral of the planar curve: $I_{\sT} =
I_\T$. This follows directly from the conservation of the geodesic
curvature $\kappa(t)$ by the mapping.

Now, \emph{if} the path on the ball is closed, then this integral is
an index. For example, a path that
does not self-intersect is of 0-index, $I_\T=0$, while  each self-intersection adds a phase of $\Delta\psi = \pm 2\pi$ (with the sign reflecting left/right handedness), or $\pm 1$ to the index. Therefore, for the sake of simplicity, in the remainder of the paper, we consider enclosed areas modulo $2\pi r^2$ and indices modulo $1$.

By the Gauss-Bonnet theorem, (see, e.g., the book \cite{kobayashi}) the surface enclosed by the path on the ball is 
\begin{equation*}
    S(r)=S=2\pi r^2 (1-{I}_{\sT})~.
\end{equation*}
Therefore \emph{if} we require $S=2\pi r^2$, then $\sT$ is of 0 index, ${I}_{\sT}=0$, and therefore also $I_\T= \frac{1}{2\pi} \Delta \psi = 0$, implying that the orientation of the ball at the end $\Omega$ of $\T$ is the same as at the beginning $\A$.

\begin{figure}
      \includegraphics[width= \columnwidth]{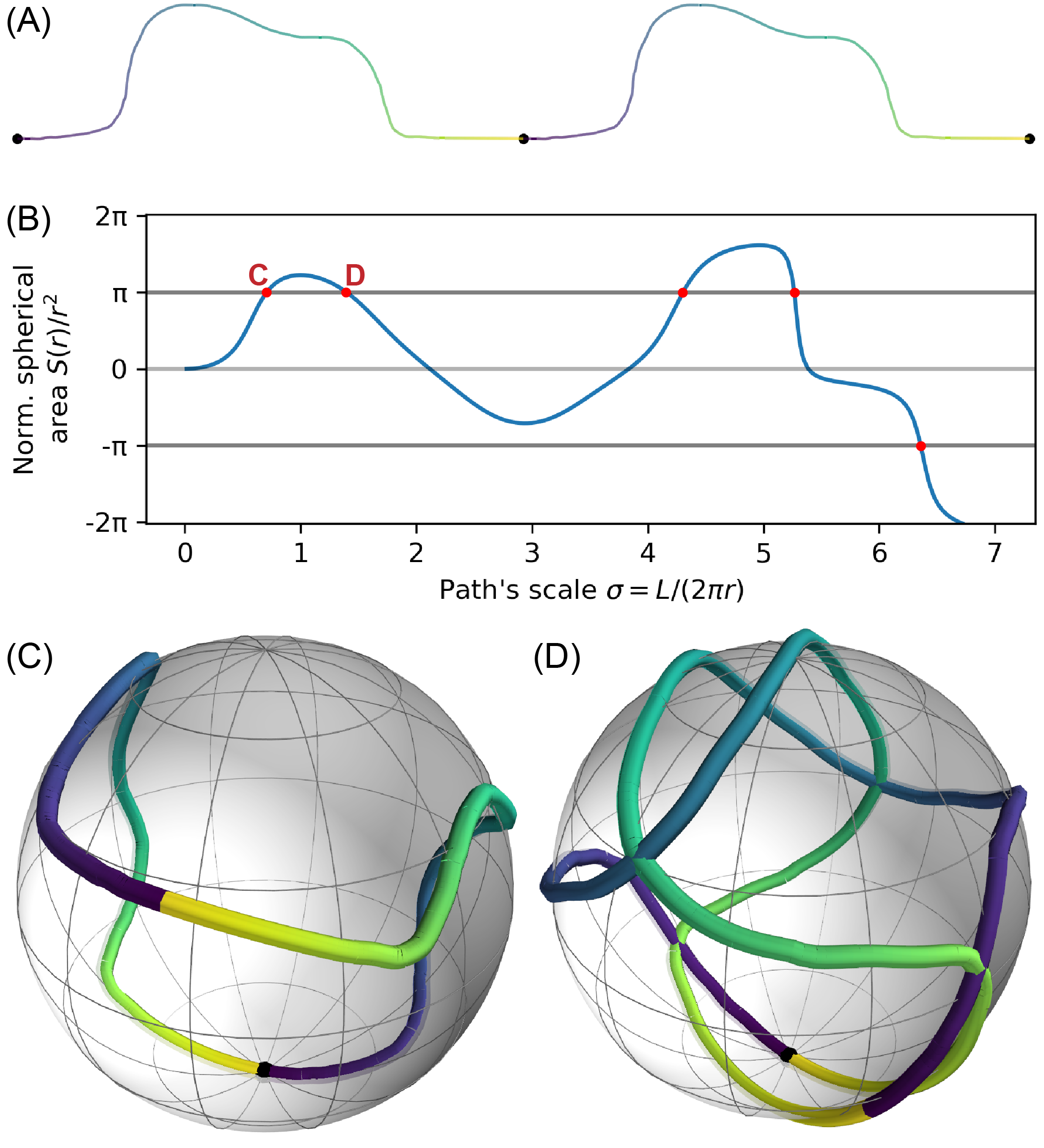}
  \caption{An illustration of the existence of multiple solutions
    i.e., radii $r$, for the path shown in (A). The area enclosed by
    the trace of a single period (green in Fig. 7) and the geodesic arc
    connecting its ends (red in Fig. 7) is plotted against
    $\sigma={L}/(2\pi r)$ on the horizontal axis. Here, $L$ is the
    length of single period of the path, $r$ is the radius of the
    ball. Five two-period trajectoid solutions are marked by red
    dots in (B). Two-period spherical traces for the first two
    solutions (at $\sigma \approx 0.707$ and at $\sigma \approx
    1.398$) are shown in (C) and (D), respectively. Color coding in
    (C) and (D) is the same as in (A).} 
  \label{fig:prince_multiple}
\end{figure}

We see that \emph{a trajectoid exists if the curve drawn on the ball encloses exactly half of its surface}, namely $2\pi r^2$. Note that for a given curve, there can be many radii $r$ for which this condition is fulfilled, as shown in \fref{fig:prince_multiple}.

Before we discuss which paths have a trajectoid that fulfills the area condition, we generalize the problem somewhat.

\section{$n-$paths}

\begin{definition}
  A path $\T$ is called a $n$-path if any trajectoid for it
  reaches a pure translation after rolling over exactly $n$ copies of $\T$, and never reaches pure translation when rolling over any smaller number $k < n$ copies of $\T$ (where both numbers are natural, $k,n \in \mathbb{N}$).
\end{definition}

The ``experimental'' situation seems to be as follows: In general, it seems
that most path are \emph{not} $1$-paths.
In other words, in this case there is no object that can recover its original
orientation after tracing just 1 copy of the (irreducible)
path. However, in stark contrast, it seems that ``most'' paths seem to
be  2-paths, although we have no proof nor a good formulation for that (see the discussion in the next section). On the other hand, the following two theorems hold:

\begin{theorem}\label{thm:1}Every  path $\T$ is an $n$-path for some finite $n$.\end{theorem}

In other words, for any path $\T$ one can construct a trajectoid which
will recover it orientation after passing trough $n$ copies of $\T$,
for some \emph{finite} $n$.

\noindent Let $C^{1,\beta }$ denote the space of differentiable
functions $\gamma:[0,1]\to \real^2$ with the derivative $\dot
\gamma\in C^\beta $.
By any path, we think of a finite $C^{1,\beta }$ path in $\real^2$ with $\beta >0$. The path may also contain corners and polygons, which may require some trivial reparameterization.

\begin{figure}[h!]
  \centering\includegraphics[width=0.9\columnwidth]{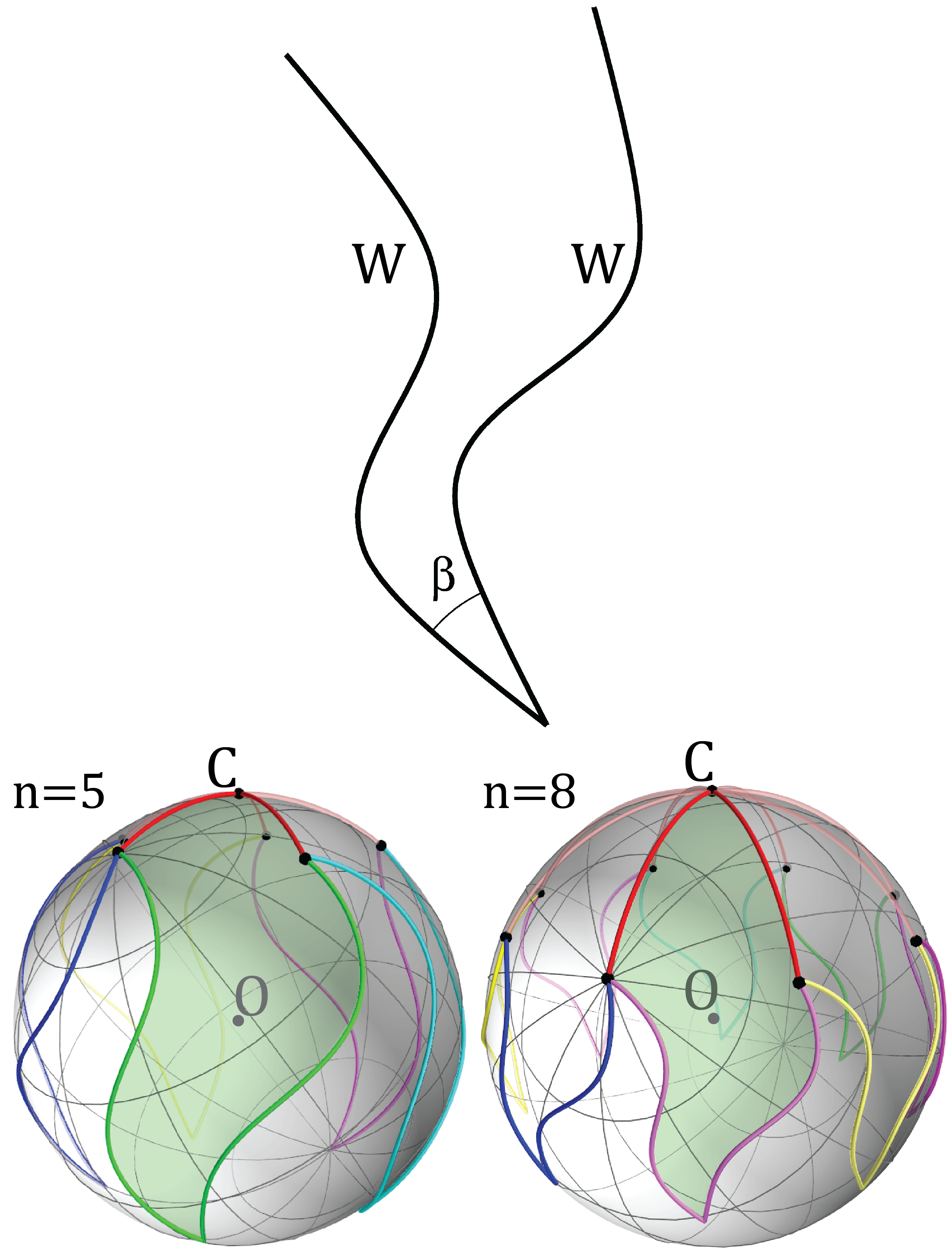}\
  \caption{A  family of paths $\T$ that are made of two identical pieces $\W$, where the second $W$ is rotated by an angle $\beta$ with respect to the first. For small enough $\beta $, $\T$ is an $n$-path with large $n > \pi/\beta$, unless $\W$ is a 1-path. Two balls below show spherical traces for trajectoid solutions for $\T$ with $n=5$ and $n=8$. In both cases, a fixed $\W$ and $\beta=\pi/4$ were used. The enclosed shaded green areas are $2\pi r^2/5$ and $2\pi r^2/8$, respectively.}
  \label{fig:wedges}
\end{figure}

On the other hand, for arbitrary finite $n$ one can easily construct
paths which are $j$-paths with $j>n$ as we show next:

\begin{theorem}\label{thm:3} Consider the path $\T$ of \fref{fig:wedges}
  which consists of a subpath $\W$ and its copy rotated by
  some angle $0 < \beta < \pi/j$, where $j \in \mathbb{N}$ and $j \ge
  2$. If $\W$ is not a $1$-path, and rolling over $\W$ is also for no $r$ a pure rotation of the ball in the axis perpendicular to the plane of the curve, then $\T$ is an $n$-path with $n>j$.
\end{theorem}

This means that for any $j$  there are indeed paths for which any
trajectoid must pass over more than $j$ copies to recover its original
orientation.

\begin{proof}[Proof of \tref{thm:1}] We consider one traversal of a
  path $\T$. After the traversal, the rotation matrix can be written
  (for each fixed $\sigma$) in an adapted coordinate system as
  \begin{equ}
R(\sigma)=  \begin{pmatrix}
    \cos{\varphi(\sigma)} & -\sin{\varphi(\sigma)} & 0\\
    \sin{\varphi(\sigma)} & \phantom{-}\cos{\varphi(\sigma)} & 0\\
     0 & \phantom{-} 0 & 1
  \end{pmatrix}~,
  \end{equ}
  where the rotation is around the $z$-axis by an angle $\varphi(\sigma)$. The rotation angle is a function of the scaled inverse radius, $\sigma=L/(2\pi r)$, with $L$ the arclength of the path.
To prove \tref{thm:1}, it suffices to note that $\phi(0)=0$, that the
trace of the rotation matrix $\tr(R(\sigma)) = 1 + 2
\cos{\varphi(\sigma)}$ equals 3 for $\varphi(\sigma)=0 \mod 2\pi$, and
that this trace is continuous in $\sigma$. If $\tr(R(\sigma))$ is
constant in $\sigma$, then $\T$ is a 1-path: The trace is equal to 3
for all $\sigma$, and therefore every $\sigma$ yields a trajectoid
that reaches a pure translation after rolling over 1 period. We do not
know how to characterize 1-paths. If $\tr(R(\sigma))$ is not constant in $\sigma$, then 
there is a $\sigma$ and a smallest $n$ for which $\tr(R(\sigma))=1+2\cos(2\pi/n)$. 
In other words, for this $\sigma$, $R(\sigma)$ is a rotation by
$\frac{2\pi}{n}$, and therefore $R(\sigma)^n$ is the identity. We see
that this path is an $n$-path. 
\end{proof}

\begin{proof}[Proof of \tref{thm:3}]

Let $R(\W,r)$ be the rotation matrix after rolling over the left part
of the path $\W$ of \fref{fig:wedges}.   Then the full rotation matrix is
$R(\T, r) = B^{-1}R^{-1}(\W, r) B \, R(\W,r)$, where
$B$ is the rotation matrix of turning by the angle $\beta$ around an axis $\nb_1 = \hat{z}$ perpendicular to the plane (the unit vector perpendicular to the plane of the drawing in \fref{fig:wedges}).  
The matrix $\tilde{B} \equiv R^{-1}(\W, r)BR(\W,r)$ is simply a representation of the matrix $B$ in a rotated coordinate system (where $R(\W,r)$ is the transformation). Therefore, the transformed $\tilde{B}$ remains a rotation by the same angle $\beta$, but around the rotated axis, the unit vector
$\nb_2(\W,r) = R(\W,r) \nb_1 $, which is different from $\nb_1$ unless $R(\W,r) = 1$.
The cosine of the net rotation angle $\gamma$ corresponding to the overall rotation, $R(\T,r)=B^{-1}\tilde{B}$, is 
\begin{equ}
  \cos{\gamma} = \cos^{2}{\beta} + \l(\nb_1
\cdot \nb_2(\W, r)\r) \sin^{2}{\beta}~,  
\end{equ}
(usually called the Rodrigues
formula, \cite{altman}[equ (19)], with some nice relations to quaternions).
Because for the unit vectors $\nb_1$, $\nb_2$, one has $|\nb_1 \cdot \nb_2| \le 1$, the value of $\cos{\gamma}$
satisfies
$$
\cos^{2}{\beta} - \sin^{2}{\beta} = \cos{2\beta}\le \cos \gamma \le 1~.
$$

Since $\beta < \pi/2$ by assumption, these bounds can be rewritten as
$0 \le \gamma \le 2\beta$. Here, $\gamma$ is considered on the
interval $[0, \pi]$ 
without any loss of generality. The path $\T$ is a 1-path if $\gamma = 0$ for some finite $r$. But this occurs only if $\nb_2(\W, r)= R(\W,r)\nb_1 = \nb_1$, and therefore $R(\W,r)$ is a rotation around vertical axis $\nb_1$, thereby contradicting the theorem's assumption about the $\W$ path.
Otherwise, the path $\T$ is a $k$-path if and only if $\gamma = 2\pi/k$, which is only possible if $2\pi/k \le 2\beta$ due to bounds above. But by assumption $\beta < \pi/j$, and therefore $\T$ can only be an $n$-path for $n>j$.
\end{proof}

\begin{figure}
\centering  \includegraphics[angle=0,width=0.6\columnwidth]{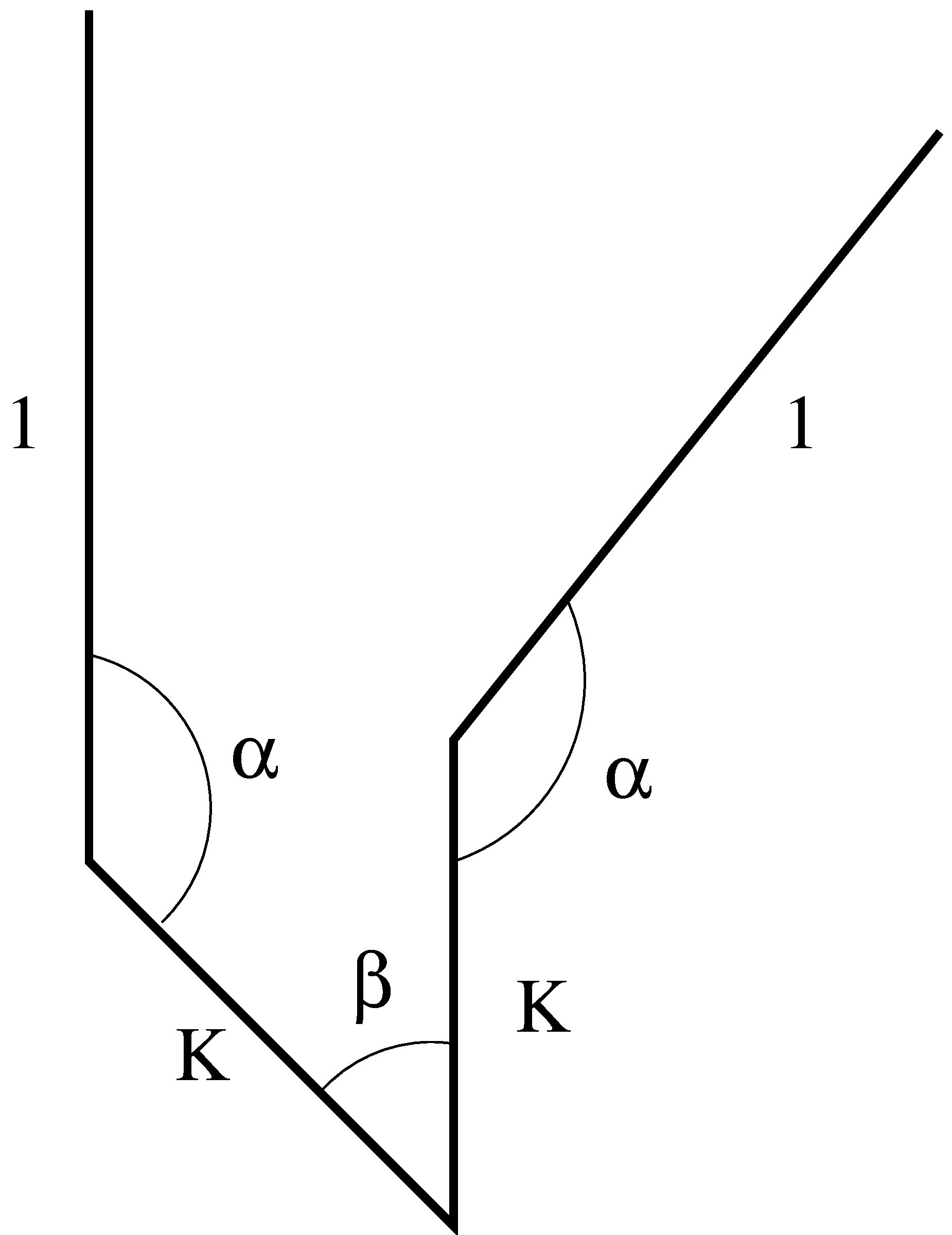}  
  \caption{A $\beta$-dependent family of paths: For each $n\ge3$ there
    is a $\beta_n$ so that with $\beta=\beta_n$ the path is an
    $n$-path.}
  \label{fig:zigzag}
\end{figure}

While \tref{thm:3} is quite general, it is perhaps a little
unsatisfactory, as it does not guarantee that for any $n\ge1$ there is a
path which is an $n$-path. We have provided such an example for an
explicit choice, as shown in \fref{fig:zigzag}. In this case, one can
show, by an explicit multiplication of the 4 rotation matrices of the
4 segments of the path, that for every $n\ge3$, there is a $\beta=\beta _n$
for which that path is an $n$-path.  The calculation was done for
K$=1/\sqrt{2}$ and $\alpha =3\pi/4$. It is important that K is
irrational, as otherwise, the left part of the figure is a 1-path,
which we also had to exclude in the proof of \tref{thm:3}.

\section{Prevalence of $2$-paths}\label{2traj}

\begin{figure}[t]
        \centering 
        \includegraphics[width= 0.8\columnwidth]{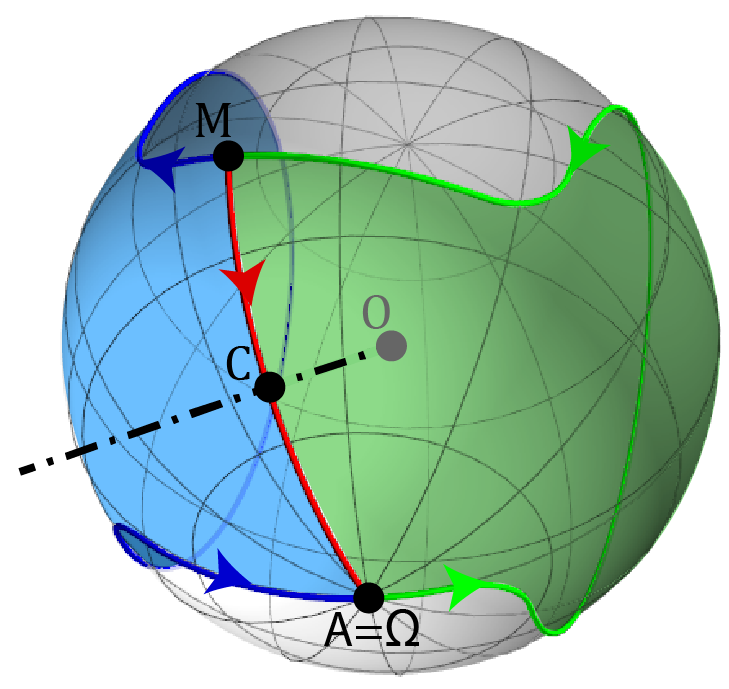}
        \caption{Adding up the two areas of $\pi r^2$ (green and blue) by passing
          through 2 copies of the original (primitive) path leads to
          the desired area $2\pi r^2$, allowing the construction of a trajectoid. Adapted with permission from Fig.~3e of \cite{xxx}. }
\label{fig:TPT}
\end{figure}

We come back to the Gauss-Bonnet theorem, and will explain, but not be
able to prove, why almost any planar curve is a $2$-path
(what ``almost'' means is discussed later). Recall that any
trajectoid is related to a closed curve on the ball, enclosing half
of its surface, i.e., $2\pi r^2$. This is difficult to obtain with
a primitive path, (a path which is not a multiple of some shorter one)
but now consider 2-paths (the following
discussion is adapted from \cite{xxx}), and let us examine what
happens after having crossed just \emph{one} copy of the path $\T$,
the green curve in  \fref{fig:TPT}.

This curve is \emph{not} closed, but we close it by the red geodesic arc between M and A. (This shortest arc
is unique unless $\M$ and $\Omega$ are antipodal, in which case a slightly
modified argument applies.)
Consider then the quantity $S(\T,r)$ corresponding to the green
area of \fref{fig:TPT}. In
other words, we call $S_{\rm g}$ is the area spanned by the (green) curve, together
with the red geodesic connecting $\M$ with $\Omega$.

It turns out experimentally, that, for almost all polygonal paths $\T$, we can find a radius
$r_*$
for which the enclosed area is either zero, or exactly $S(\T,r_*) = \pi r_*^2$. Assuming that
we found such an $r_*$ then by the Gauss-Bonnet theorem the index of
the closed green + red curve equals one half, namely $I_{\rm g} = \frac{1}{2
  \pi}\int{\d t \, \kappa(t)} =  1 - S_{\rm g}/(2 \pi  r^2) =
\frac{1}{2}$. However, the integral of the curvature vanishes along
the great arc (because it is a geodesic), and along $\sT= \sAM$ (due
to the periodicity of the planar curve $\T$, $\Delta \psi = 0$). It
follows, that the only contribution to the index comes from the
corners $A$ and $M$ whose angles add up to $2 \pi I_{\rm g}
= \pi$. 

This facilitates the following construction, using \fref{fig:TPT}: we rotate the\ green curve
by \SI{180}{\degree} about the mid-point of the red arc to form the
blue curve. Because the two angles add up to $\pi$, the blue and green
curves are connected \emph{smoothly}, that is, without corners. The
connected curve is therefore the mapping to the ball of a two-period
repeat  of $\T$. Now, since the green area is $S_{\rm g} = \pi r^2$
and is equal to the blue area $S_{\rm b}$, the two areas add up to
$2\pi r^2$. Thus, by doubling the non-closed path, we achieved the
condition that the enclosed area is that of a half-ball, and hence a
trajectoid exists for $\T$. It is easy to see that the same argument
applies also to paths that begin at a sharp corner, simply by shifting the starting point to a smooth point on the curve. 

But, we still have to find an $r$ for which the green area equals $\pi
r^2$. We therefore consider the normalized green (or blue) area $S(\T,r)/r^2$ as a function of given path $\T$ and a ball of radius $r$. Unfortunately, the
function $S(\T,r)/r^2$ does not have nice monotonicity properties. (In
\fref{fig:prince_multiple} we use the more natural variable
$\sigma=L/(2\pi r)$ for the dependence on the radius.) On the other
hand, it proves relatively difficult to construct paths for which
$S(\T,r)/r^2\ne n\pi, n \in \mathbb{Z}$ for all $r$, as we have seen
in \tref{thm:3}.

So at this point, we are left with the empirical result that for ``most'' paths $\T$ we indeed find an $r$ for which a trajectoid exists when one passes \emph{two} copies of the path $\T$. Numerical experimentation shows that for randomly chosen paths $\T$, one always seems to find an $r$ for which $S(\T,r)/r^2=\pi$. This motivates us to formulate this as a
conjecture, but the reader should be aware that the space of finite
curves on the plane is infinite-dimensional. In such contexts, notions like measure or genericity are delicate, and it can happen that sets of full measure only contain ``uninteresting'' examples.

A possible way out could be the piecewise affine interpolation of the curve, which is perhaps closest to the way trajectoids are constructed.
Another possibility is to work in Fourier space, imposing conditions on the Fourier coefficients, see e.g., \cite{kahane}.

Let $C^{1,\beta }$ denote the space of differentiable functions $\gamma:[0,1]\to \real^2$ with the derivative $\dot \gamma\in C^\beta $.  
\begin{conjecture}
The subset of piecewise $C^{1,\beta} $, $\beta\in(0,1]$ curves
    which are $2$-paths,
i.e., for which there is an $r$ where the enclosed area satisfies $S(P,r)=\pi r^2$, is \textbf{dense} in $C^{1,\beta}$.
\end{conjecture}
The ``piecewise'' condition can be omitted if one allows for reparameterizations of the $\gamma$.

\begin{figure}[t]
        \centering 
        \includegraphics[width=\columnwidth]{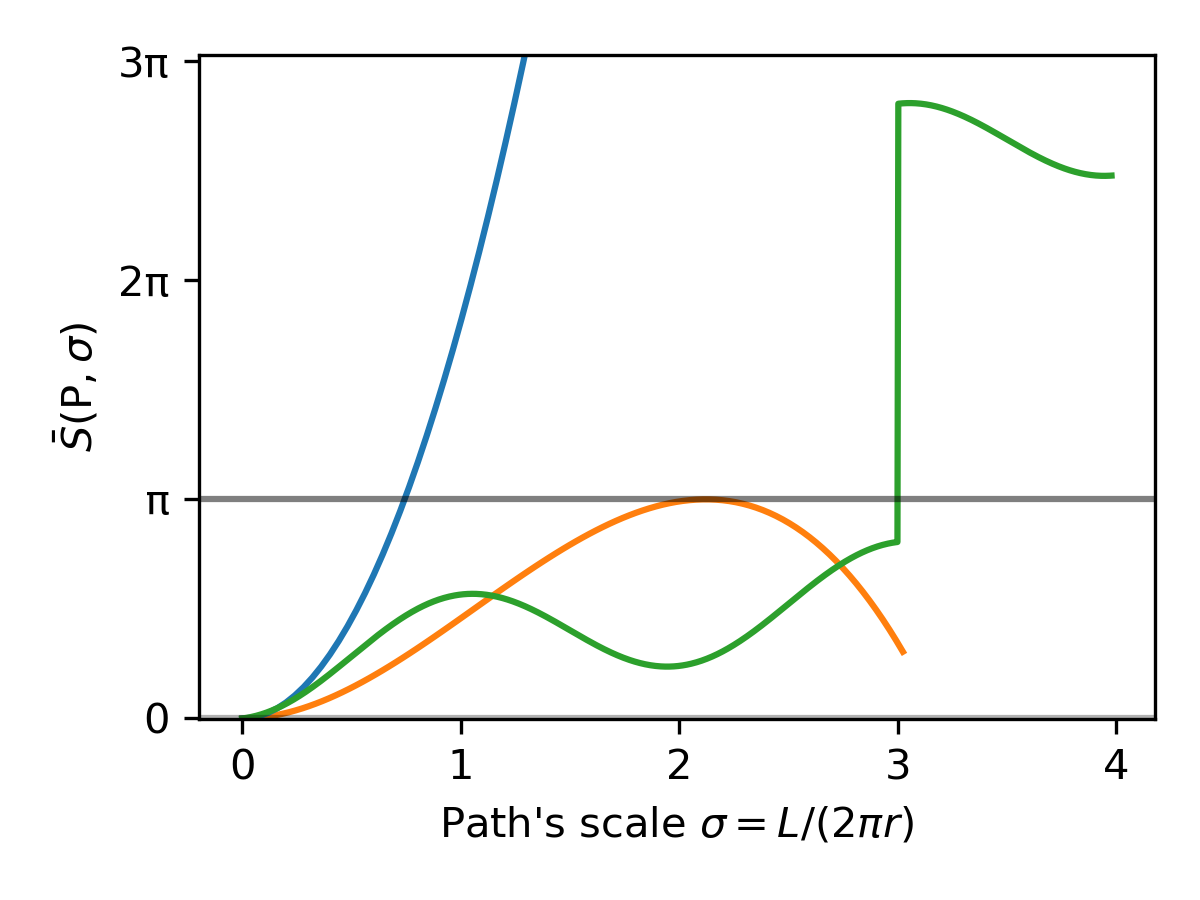}
        \caption{The normalized enclosed area function 
          $\bar{S}(\T,\sigma)$, for three choices of $\T$. The generic case is transversal crossing of $\pi$ (blue), whereas the non-generic case are tangency to $\pi$ (orange) and a jump by $2\pi$ caused by antipodality of $\A$ and $\Omega$ (green)}
\label{fig:generic}
\end{figure}

Still we can add the following density result: Assume that some path $\T$
is a $2$-path, i.e.,
$S(\T,r_*)/r_*^2=\pi\mod 2\pi$.
In general, all paths in a small open $C^2$ neighborhood
of $\T$ are again $2$-paths. Indeed, if the function defined as
\begin{equation*}
        \bar{S}\l(\T,\sigma \r)\equiv \frac{S\l(\T,r\r)}{r^2} \Big|_{r = L/(2\pi \sigma)}
\end{equation*}
traverses $\pi\mod 2\pi$ transversely -- which 
is the generic case -- then, by continuity of the area, for curves near $\T$
that transversality is maintained and we have a $2$-path. One sees
that the 2-path property is mostly an open condition in the
space of paths. On the other hand, we conjecture above that close to
any non-2-path, there is one which leads to a 2-path.

We illustrate the typical generic and non-generic cases in
\fref{fig:generic}. The generic situation appears when
$\bar{S}(\T,\sigma)$ crosses the level $\pi$
transversally. Two notable non-generic cases appear if $\bar{S}(\T,\sigma)$
is tangent to the level $\pi$, or if $\bar{S}(\T,\sigma)$ jumps because for
some $\sigma^*$ the points $\A$ and $\Omega$ are antipodal.

Note that, for fixed $\beta _n$, the function $\tr(\tilde S(\beta_n ,\sigma  ))$
oscillates (irregularly) in $[3-2\beta_n ^2,3]$ as a function of $\sigma $,
and so, we can indeed see (cf.~\fref{fig:prince_multiple}) that in general, there can be many
trajectoids for a given curve (usually with decreasing radii $r$). It
seems that the trajectoid with largest $r$ will roll most accurately
along its path.

 \section{Fabricating a trajectoid and further remarks}\label{sec:details}

 \begin{figure}[h]
        \centering 
        \includegraphics[angle=0,width= 0.9 \columnwidth]{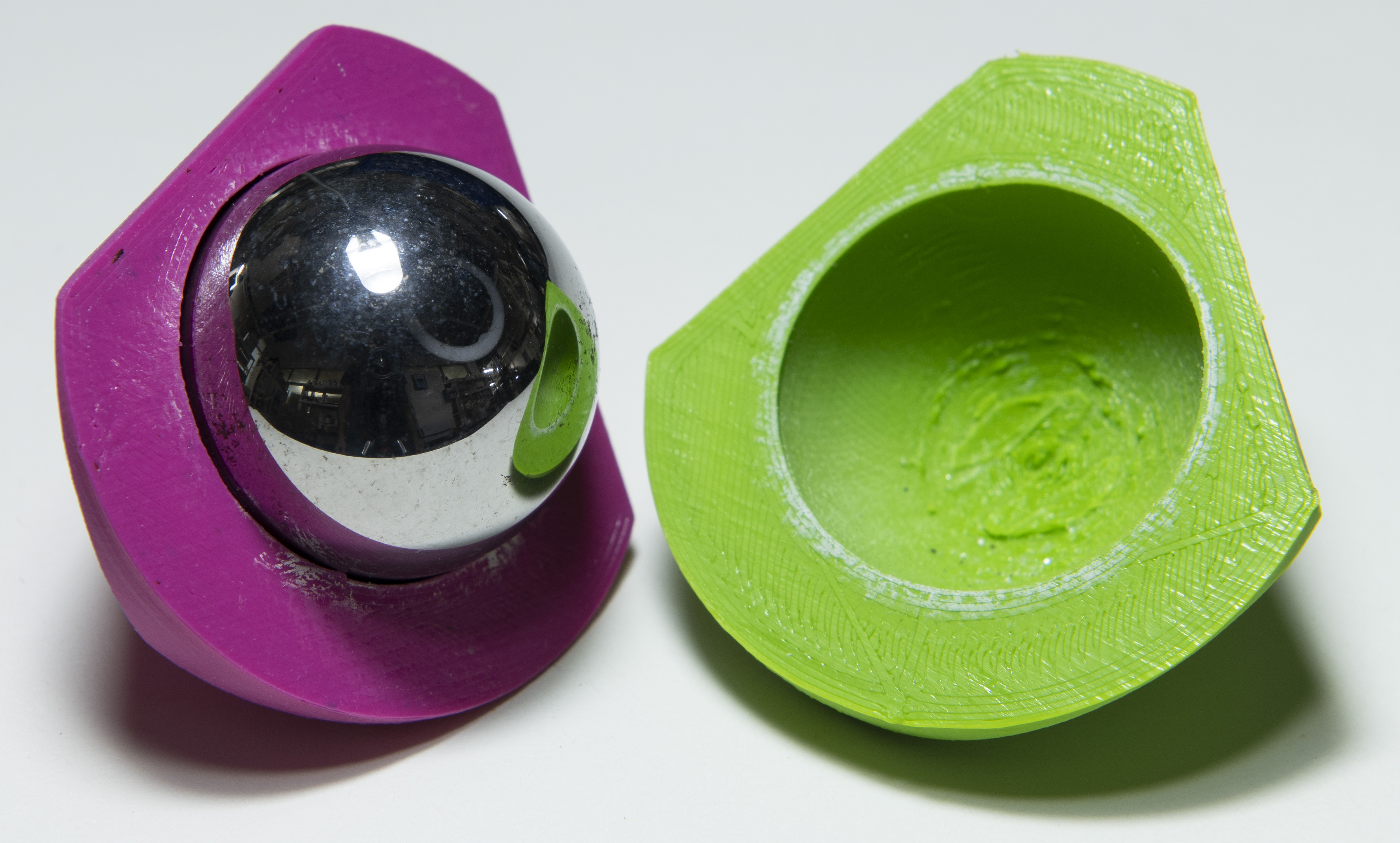}
        \caption{An example of the fabrication of the two-period trajectoid from \fref{fig:rollingprince}. The steel ball is 6.24 times more dense than the surrounding plastic (PLA) shell made of two identically shaped halves (magenta and green) 3D-printed and then glued together.}
\label{fig:fabrication}
\end{figure}
\begin{enumerate}
 \item Readers can fabricate a trajectoid for a given curve of their
   choice in the following way. Prepare your curve as two columns of
   (x,y) coordinates in a .csv file (comma-separated values) and load
   it into the online tool (Google Colab notebook, link below) and
   follow the instructions therein to obtain 3D-printer-ready files
   (.stl) for a trajectoid of your curve. For the quite common case of
   a trajectoid for a 2-path, you 3D-print two .stl
   files (one for each half). insert the steel ball into the core
   cavity and glue the two halves together. The typical quality of the pieces we have tested  is illustrated in \fref{fig:rollingprince}. The gluing together of the steel ball and the `weightless' outer blue piece--which is made by a 3D-printer--is shown \fref{fig:fabrication}.
  A link to a Google Colab notebook can be found on 
   \href{https://github.com/yaroslavsobolev/trajectoids}{https://github.com/yaroslavsobolev/trajectoids}. We
   used stainless steel balls of diameter 1 in, weight 66.7g.

\item Because of the half-surface condition mentioned above, it
  turns out that the trajectoid for a $2$-path is actually made of two identical pieces as shown in \fref{fig:rollingprince} and \fref{fig:fabrication}.
\item One can approximate any smooth curve by finer and finer
  polygons, shaving off each time a tiny piece for each infinitesimal segment.
  \item We do not know precisely which primitive paths are 1-paths
    (``primitive" excludes paths that are just concatenations of two
    (or more) identical pieces). But any path which is a ``V'' (e.g.,
    the lines connecting  $(-x,y),(0,0),(x,y)$ with $x>0$ and $y>0$)
    is a $1$-path. The problem probably needs a careful analysis of symmetries of the paths.

\end{enumerate}

\noindent{\bf{Acknowledgements}}: We thank the referees for their careful reading of the manuscript and pointing
out a number of inconsistencies in the original text. JPE is partially supported by SwissMap. YS and TT are supported by the Institute for Basic Science, Republic of
Korea, project code IBS-R020-D1.

\bibliography{notices}
\end{document}